\documentclass[12pt]{iopart}
\usepackage{iopams}

\begin{document}

\title[Non-distributive algebraic structures]%
{Non-distributive algebraic structures 
derived from nonextensive statistical mechanics}

\author{
Pedro G. S. Cardoso$^1$, 
Ernesto P. Borges$^2$, 
\\ Thierry C. P. Lob\~ao$^3$, 
Suani T. R. Pinho$^1$
}

\address{$^1$Instituto de F\'{\i}sica, Universidade Federal da Bahia \\
Campus Universit\'ario de Ondina, 40210-340 Salvador--BA, Brazil}
\address{$^2$Escola Polit\'ecnica, Universidade Federal da Bahia \\
Rua Prof. Aristides Novis 2, 40210-630 Salvador--BA, Brazil}
\address{$^3$Instituto de Matem\'atica, Universidade Federal da Bahia \\
Campus Universit\'ario de Ondina, 40170-110 Salvador--BA, Brazil}

\ead{peufisbach@gmail.com, ernesto@ufba.br, thierry@ufba.br, suani@ufba.br}


\begin{abstract}

We propose a two-parametric non-distributive algebraic structure that
follows from $(q,q')$-logarithm and $(q,q')$-exponential functions.
Properties of generalized $(q,q')$-operators are analyzed.
We also generalize the proposal into a multi-parametric structure
(generalization of logarithm and exponential functions and the
corresponding algebraic operators).
All $n$-parameter expressions recover $(n-1)$-generalization when
the corresponding $q_n\to1$.
Nonextensive statistical mechanics has been the source of successive 
generalizations of entropic forms and mathematical structures, 
in which this work is a consequence.
\newline

\noindent
{\bf Keywords:} Nonextensive statistical mechanics, deformed functions,
deformed algebraic structures
\newline

\end{abstract}


\section{Introduction}

The $q$-logarithm and the $q$-exponential functions \cite{ct:quimicanova}
appears in the very foundations of nonextensive statistical mechanics
\cite{ct:1988},
and they are present in (virtually) every development of the theory and
its applications.
These functions are generalizations of the usual logarithm and exponential,
given by
\begin{equation}
 \label{q-log}
 \ln_q x := \frac{x^{1-q}-1}{1-q}, \quad x>0,
\end{equation}
and
\begin{equation}
 \label{q-exp}
 e_q^x := [1+(1-q)x]_+^{\frac{1}{1-q}},
\end{equation}
where $[p]_+ := \max\{p,0\}$.
They are inverse of each other, and recover their usual counterparts as $q\to1$.
The $q$-logarithm function allows the nonextensive entropy be neatly written
as
\begin{equation}
 S_q = k \sum_{i=1}^W p_i \ln_q (1/p_i),
\end{equation}
and also the $q$-canonical distribution is given by a $q$-exponential function,
making a close formal parallelism with Boltzmann-Gibbs
statistical mechanics.

The $q$-logarithm and $q$-exponential functions have led to the development
of a $q$-algebra \cite{wang:q-algebra,epb:q-algebra}, by the definition of
a $q$-sum
\begin{equation}
 \label{q-sum}
 x \oplus_q y := x + y + (1-q) xy
\end{equation}
and a $q$-product
\begin{equation}
 \label{q-product}
 x \otimes_q y := \left[ x^{1-q} + y^{1-q} -1 \right]_+^{\frac{1}{1-q}},
 \quad x>0,\, y>0.
\end{equation}

These operators allow to write properties of $\ln_q x$ and $e_q^x$ in a
simple and compact form, {\em e.g.},
\begin{equation}
 \label{lnq(xy)}
 \ln_q (xy) = \ln_q x \oplus_q \ln_q y,
\end{equation}
\begin{equation}
 \label{lnq(xotimesqy)}
 \ln_q (x \otimes_q y) = \ln_q x + \ln_q y,
\end{equation}
and correspondingly,
\begin{equation}
 \label{eq(xy)}
 e_q^x e_q^y = e_q^{x \oplus_q y},
\end{equation}
\begin{equation}
 \label{eq(x+y)}
 e_q^{x+y} = e_q^x \otimes_q e_q^y.
\end{equation}

It is possible to define the inverse operators $q$-difference and $q$-ratio
(and $q$-power, $q$-root), yielding a non-distributive $q$-algebraic structure.

There have been proposals of further generalizations of either $S_q$ or
$q$-logarithm and $q$-exponential functions (or both), by means of two
parameters.
One instance was proposed by Papa \cite{papa}, 
who developed an infinite set of entropies $_nS_q$
($n= 1, 2, \dots$ is called order of the entropy).
$S_q$ entropy is recovered with $n=1$,
and Boltzmann-Gibbs entropy is recovered by either $q\to1$ or $n\to\infty$.

A two-parameter ($q$,$q'$)-entropy $S_{q,q'}$ was proposed by Roditi and
one of us \cite{epb-roditi}; it is based on Chakrabarti and Jagannathan
two-parameter derivative \cite{chakrabarti-jagannathan},
following Abe's rule \cite{abe} for generating entropies.
This $S_{q,q'}$ entropy is symmetric by exchange of indexes
$q \leftrightarrow q'$
and recovers nonextensive entropy $S_q$ if either $q$ or $q'$ is set to unity.
From $S_{q,q'}$, the authors define a two-parameter ($q$,$q'$)-logarithm ---
its inverse, the ($q$,$q'$)-exponential, has an implicit form.
$S_{q,q'}$ is, in fact, a simple variation of Mittal two-parameter entropy
\cite{mittal}, proposed years before the nonextensive seminal paper
\cite{ct:1988}, as remarked by \cite{kaniadakis:pre2005}.

Kaniadakis and co-authors \cite{kaniadakis:pre2005} propose a different
two-parameter entropy, not immediately connected to nonextensive $S_q$,
and coherently introduce a two-parameter
($\kappa$,$r$)-deformed logarithm and exponential functions.
Their work is a consequence of previous papers, starting with
\cite{kaniadakis:physa2001}, in which it is introduced the $\kappa$-exponential,
that satisfies $\exp_{\kappa}^{-1} (x) = \exp_{\kappa}(-x)$
(differently from the $q$-exponential given by Eq.~(\ref{q-exp}), which
obeys $e_q^{-1}(x)=e_q^{(-x/(1+(1-q)x))}$).

Along a different path, Tsallis, Bemski and Mendes \cite{ct-bemski-mendes}
introduced a two-parameter generalization of the $q$-exponential,
Eq.~(\ref{q-exp}):
the successive generalization of the exponential function
($e^x$, $e_q^x$, $e_{q,1}^x$, $e_{q,q'}^x$)
is obtained by the generalization of the differential equation it obeys.
Their ($q$,$q'$)-exponential is presented as a function of hypergeometric
functions (it cannot be expressed explicitly), and recover the one-parameter
$e_q^x$ as $q=q'$. This ($q$,$q'$)-exponential presents two power law regimes,
with a crossover between them (as a particular case, with $q'=1$,
it can exhibit a crossover from a power-law regime to an exponential tail),
and this interesting feature has been observed in a variety of phenomena
\cite{ct-bemski-mendes,montemurro,epb:musicians,ct-anjos-epb,epb:gdp}.

Recently a still different proposal have appeared in literature:
Schw\"ammle and Tsallis \cite{schwammle-ct} introduce a two-parameter
generalization of the logarithm and exponential functions.
The ($q$,$q'$)-logarithm is properly defined in order to satisfy
\begin{equation}
 \label{veit-ct}
 \ln_{q,q'} (x \otimes_{q} y) = \ln_{q,q'} x \oplus_{q'} \ln_{q,q'} y,
\end{equation}
which is a generalization of two basic properties of the one-parameter
$q$-logarithm, Eq.~(\ref{lnq(xy)})--(\ref{lnq(xotimesqy)}),
and, correspondingly, the ($q$,$q'$)-exponential satisfies
\begin{equation}
 e_{q,q'}^{x\oplus_{q'} y} = e_{q,q'}^x \otimes_{q} e_{q,q'}^y,
\end{equation}
that generalizes Eq.~(\ref{eq(xy)})--(\ref{eq(x+y)}).

In our present work, we use Schw\"ammle and Tsallis ($q$,$q'$)-deformed
functions and define two-parameter algebraic operators,
following the lines of \cite{kaniadakis:pre2005}
and generalizing \cite{wang:q-algebra,epb:q-algebra}.

It is usual that different forms of deformed functions receive the same notation
and are called by similar names, $q$-exponential, ($q$,$q'$)-exponential,
and so on; but this shall not confound the reader, and (s)he must be able
to distinguish between them by the context.

The paper is organized as follows:
Sec. \ref{sec:qqprime-algebra} introduces a new two-parameter algebra and
presents some of its properties;
Sec. \ref{sec:qqprime-dot_product} proposes a different two-parameter
deformation of the product;
multi-parametric generalization of functions and algebras are addressed
in Sec. \ref{sec:multiparametric}. Finally, in Sec. \ref{sec:conclusions}
we draw our conclusions and present some perspectives.
The relative simplicity of the two-parameter equations, when expressed
by means of the $q$-deformed versions, as they appear in the main body
of the paper, contrast with the explicit forms (presented in the Appendix)
that are not so simple and clean.

\section{A Two-Parameter Algebra}
\label{sec:qqprime-algebra}

Let us rewrite the basic properties
$e^{x+y}=e^x e^y$ and $\ln (xy) = \ln x + \ln y$
as
\begin{equation}
 \label{recipe-0}
 \begin{array}{lrc}
  x+y &\equiv& \ln(e^x e^y),
  \\ \noalign{\medskip}
  xy  &\equiv& e^{\ln x + \ln y}.
 \end{array}
\end{equation}
We may take these expressions as a basis for a redefinition of the
$q$-algebraic operators:
the $q$-sum (Eq.~(\ref{q-sum})) and the $q$-product (Eq.~(\ref{q-product}))
can be found by the following recipe:
replace the logarithm and exponential of Eq.~(\ref{recipe-0})
by their $q$-generalizations:
\begin{equation}
 \label{recipe-1}
 \begin{array}{rcl}
  x\oplus_q y  &:=& \ln_q(e_q^x\,e_q^y),
  \\ \noalign{\medskip}
  x\otimes_q y &:=& e_q^{\ln_qx+\ln_qy}.
 \end{array}
\end{equation}

In Ref. \cite{kaniadakis:pre2002}, where this recipe was proposed,
it was shown that other pairs of function--inverse function can be used
(and not only the logarithm--exponential pair),
and thus yielding different deformed algebraic operators.
We shall apply this recipe in the following.

The two-parameter generalization of the $q$-logarithm and
$q$-exponential functions introduced in \cite{schwammle-ct} can be
written as (see Eq.~(\ref{app-log}) and (\ref{app-exp}) in the Appendix 
for the explicit forms)
\begin{equation}
 \label{qq'-logarithm}
 \ln_{q,q'}x = \ln_{q'}e^{\ln_qx},
\end{equation}
\begin{equation}
 \label{qq'-exponential}
 e_{q,q'}^x = e_q^{\ln e_{q'}^x}.
\end{equation}
Though these ($q$,$q'$)-functions are {\em not} symmetric by
interchange of parameters $q\leftrightarrow q'$, the mono-parametric
$q$-functions are recovered by either $q\to1$ or $q'\to1$. They are
inverse of each other, as may be promptly verified by
$\ln_{q,q'}e_{q,q'}^x=e_{q,q'}^{\ln_{q,q'}x}=x$.
Ref.~\cite{schwammle-ct} brings various properties of these
functions.

Here we propose a natural extension of Eq.s~(\ref{recipe-1})
for a two-parameter version:
\begin{equation}
 \label{recipe-2-sum}
 x\oplus_{q,q'} y := \ln_{q,q'}(e_{q,q'}^x\,e_{q,q'}^y)
\end{equation}
and
\begin{equation}
 \label{recipe-2-product}
 x\otimes_{q,q'} y := e_{q,q'}^{\ln_{q,q'}x+\ln_{q,q'}y}.
\end{equation}

It is possible to rewrite the ($q$,$q'$)-sum and the ($q$,$q'$)-product
expressing them by means of the (mono-parametric) $q$-functions:
\begin{equation}
 x\oplus_{q,q'}y = \ln_{q'}e^{\ln e_{q'}^x\oplus_q\ln e_{q'}^y},
\end{equation}
with $1+(1-q')x>0$, $1+(1-q')y>0$, and
\begin{equation}
 x\otimes_{q,q'}y = e_q^{\ln(e^{\ln_qx}\otimes_{q'}e^{\ln_qy})},
\end{equation}
with $x>0$, $y>0$, 
$\left[(e^{\ln_q x})^{1-q'}+(e^{\ln_q y})^{1-q'}\right]>1$.
Explicit expressions are, respectively, 
Eq.~(\ref{app-sum}) and Eq.~(\ref{app-product}).

Let us address some properties of the ($q,q'$)-sum:
it is commutative,
\begin{equation}
 x\oplus_{q,q'}y=y\oplus_{q,q'}x,
\end{equation}
associative,
\begin{equation}
 x\oplus_{q,q'}(y\oplus_{q,q'}z)=(x\oplus_{q,q'}y)\oplus_{q,q'}z,
\end{equation}
and presents neutral element:
\begin{equation}
 x\oplus_{q,q'}0=x\,,\forall q,q' \in \mathbb{R}.
\end{equation}
The opposite element is given by:
\begin{equation}
 x\oplus_{q,q'}(\ominus_{q,q'}x)=0,
\end{equation}
which implies (see also (\ref{app-opposite}))
\begin{equation}
 \label{qq'-opposite}
 \ominus_{q,q'}x=\ln_{q'}e^{\ominus_q(\ln e_{q'}^x)},
\end{equation}
with $1+(1-q')x>0$, $x\ne\ln_{q'}e^{\frac{-1}{1-q}}$.

Using Eq.~(\ref{qq'-opposite}) we can define the ($q,q'$)-difference as:
\begin{equation}
 \label{qq'-difference}
 x\ominus_{q,q'}y := x\oplus_{q,q'}(\ominus_{q,q'}y) =
 \ln_{q'}e^{\ln e_{q'}x\ominus_q\ln e_{q'}y}
\end{equation}
with  $1+(1-q')x>0$, $1+(1-q')y>0$, $y\ne\ln_{q'}e^{\frac{-1}{1-q}}$
(see Eq.~(\ref{app-difference})),
which is a generalization of the $q$-difference operator
\cite{wang:q-algebra,epb:q-algebra},
\begin{equation}
 x \ominus_q y = \frac{x-y}{1+(1-q)y},
 \quad y \ne \frac{-1}{1-q}.
\end{equation}

The ($q,q'$)-sum is non-distributive:
\begin{equation}
 a(x\oplus_{q,q'}y) \ne (ax)\oplus_{q,q'} (ay),
 \,\forall a\ne0, a\ne1,\, \forall q, q'\in \mathbb{R}.
\end{equation}

Now we present the corresponding properties of the ($q,q'$)-product:
commutativity,
\begin{equation}
 x\otimes_{q,q'}y=y\otimes_{q,q'}x,
\end{equation}
associativity,
\begin{equation}
 x\otimes_{q,q'}(y\otimes_{q,q'}z)=(x\otimes_{q,q'}y)\otimes_{q,q'}z,
\end{equation}
it presents neutral element,
\begin{equation}
 x\otimes_{q,q'}1=x\,,\forall q, q' \in \mathbb{R},
\end{equation}
and its inverse is given by
\begin{equation}
 x\otimes_{q,q'}(1\oslash_{q,q'}x) = 1,
\end{equation}
which implies (see also Eq.~(\ref{app-inverse}))
\begin{equation}
 \label{qq'-inverse}
 1\oslash_{q,q'}x = e_q^{\ln(1\oslash_{q'} e^{\ln_qx})},
\end{equation}
with $0<x<e_q^{\frac{\ln 2}{1-q'}}$ for $q'<1$, 
and $x>e_q^{-\frac{\ln 2}{q'-1}}$ for $q'>1$.

Using Eq.~(\ref{qq'-inverse}) we can define the ($q,q'$)-ratio as
(see Eq.~(\ref{app-ratio})):
\begin{equation}
 \label{qq'-ratio}
 x\oslash_{q,q'}y := x\otimes_{q,q'}(1\oslash_{q,q'}y) =
 e_q^{\ln(e^{\ln_qx}\oslash_{q'}e^{\ln_qy})},
\end{equation}
with $x>0$, $y>0$, 
$\left[(e^{\ln_q x})^{1-q'}-(e^{\ln_q y})^{1-q'}+1\right]>0$.
This equation recovers the $q$-ratio as $q'\to1$ (or $q\to1$):
\begin{equation}
 x \oslash_q y
 = [x^{1-q} - y^{1-q} + 1]_+^{\frac{1}{1-q}},
 \quad x>0, y>0.
\end{equation}

The ($q,q'$)-product is non-distributive:
\begin{equation}
 a\otimes_{q,q'}(x+y)\ne
 (a\otimes_{q,q'}x) + (a\otimes_{q,q'}y)\,,
 \forall a\ne1\,,\forall q, q' \in \mathbb{R}.
\end{equation}

The ($q$,$q'$)-logarithm and ($q$,$q'$)-exponential satisfy the following
relations,  expressed by means of these new two-parameter operators
(restrictions of ($q,q'$)-operators and ($q,q'$)-functions shall apply):
\begin{eqnarray}
\begin{array}{l@{\;}c@{\;}ll@{\;}c@{\;}l}
 \ln_{q,q'}(xy)&=&\ln_{q,q'}x\oplus_{q,q'}\ln_{q,q'}y
& \qquad
 e_{q,q'}^{x\oplus_{q,q'}y}&=&e_{q,q'}^x\,e_{q,q'}^y
\medskip \\
 \ln_{q,q'}(x\otimes_{q,q'}y)&=&\ln_{q,q'}x+\ln_{q,q'}y
& \qquad
 e_{q,q'}^{x+y}&=&e_{q,q'}^x\otimes_{q,q'}e_{q,q'}^y
\medskip \\
 \ln_{q,q'}(x/y)&=&\ln_{q,q'}x\ominus_{q,q'}\ln_{q,q'}y
& \qquad
 e_{q,q'}^{x\ominus_{q,q'}y}&=&e_{q,q'}^x/e_{q,q'}^y
\medskip \\
 \ln_{q,q'}(x\oslash_{q,q'}y)&=&\ln_{q,q'}x-\ln_{q,q'}y
& \qquad
 e_{q,q'}^{x-y}&=&e_{q,q'}^x\oslash_{q,q'}e_{q,q'}^y
\end{array}
\end{eqnarray}

These ($q$,$q'$)-operators are also {\em not} symmetric in relation to the
interchange of parameters $q\leftrightarrow q'$, and they are reduced to the
mono-parametric $q$-operators if either $q$ or $q'$ are set to unity,
similarly to the behavior exhibited by the ($q$,$q'$)-functions that
have originated them.

\section{The ($q$,$q'$)-Dot Product}
\label{sec:qqprime-dot_product}

A different deformed product, introduced in \cite{epb:q-algebra},
is defined by repeated $q$-sums of $n$ ($n\in\mathbb{N})$ equal terms:
\begin{equation}
 \label{q-dot}
 n\odot_q x :=
 \underbrace{x\oplus_qx\oplus_q\cdots\oplus_qx}_{n\,\mathrm{times}} =
 \frac{[1+(1-q)x]_+^n - 1}{1-q}.
\end{equation}
Of course we have $n\odot_{q\rightarrow1}x=nx$.
It is possible to extend its definition from
$ \odot_q : \mathbb{N} \times \mathbb{R} \rightarrow \mathbb{R}$ to
$ \odot_q : \mathbb{R} \times \mathbb{R} \rightarrow \mathbb{R}$:
\begin{equation}
 \label{q-dotR}
 a\odot_q x := \frac{[1+(1-q)x]_+^a - 1}{1-q}.
\end{equation}

This product is not commutative, however it determines a one-dimensional 
vector space structure over $\mathbb{R}$ if we observe that
\begin{equation}
 a \odot_q (x \oplus_q y)=(a \odot_q x) \oplus_q (a \odot_q y),
\end{equation}
and
\begin{equation}
 \label{pseudo-distributive}
 (a+b) \odot_q x = (a \odot_q x) \oplus_q (b \odot_q x)
\end{equation}
($a, b, x, y \in \mathbb{R}$).

The $q$-dot product may be expressed by means of the $q$-exponential
and $q$-logarithm functions:
\begin{equation}
 a \odot_q x = \ln_q[(e_q^x)^a].
\end{equation}
A interesting property of the $q$-dot product is
\begin{equation}
 e_q^x = [e_q^{(a\odot_qx)}]^{\frac{1}{a}}.
\end{equation}

We can proceed analogously and define the ($q,q'$)-dot product as:
\begin{equation}
 \label{qq'-dot}
 n \odot_{q,q'} x :=
 \underbrace{x\oplus_{q,q'}x\oplus_{q,q'}\cdots\oplus_{q,q'}x}
            _{n\,\mathrm{times}}
 \qquad (n\in \mathbb{N}),
\end{equation}
that may be expressed as
\begin{equation}
 n \odot_{q,q'} x = \ln_{q,q'} [ (e_{q,q'}^x)^n ],
\end{equation}
and extended to
$ \odot_q : \mathbb{R} \times \mathbb{R} \rightarrow \mathbb{R}$,
turning into
\begin{equation}
 \label{recipe-qq'-dot}
 a \odot_{q,q'} x = \ln_{q,q'} [ (e_{q,q'}^x)^y ]
\end{equation}
(see Eq.~(\ref{app-dot})).

Some properties of the ($q,q'$)-dot product ($a$, $x$, $y$ $\in \mathbb{R}$):
\begin{equation}
 a \odot_{q,q'} x = \ln_{q'} e^{a \odot_q \ln e_{q'}^x},
\end{equation}
\begin{equation}
 e_{q,q'}^x = [ e_{q,q'}^{(a\odot_{q,q'}x)} ]^{\frac{1}{a}},
\end{equation}
\begin{equation}
 a\odot_{q,q'}(x\oplus_{q,q'}y)=(a\odot_{q,q'}x)\oplus_{q,q'}(a\odot_{q,q'}y),
\end{equation}
\begin{equation}
 (a+b) \odot_{q,q'} x = (a\odot_{q,q'}x) \oplus_{q,q'} (b\odot_{q,q'}x).
\end{equation}

\section{Multi-parametric Functions and Algebras}
\label{sec:multiparametric}

We can further generalize logarithms and exponentials into a
multi-parametric family of functions, and correspondingly,
generate multi-parametric algebraic structures. For instance,
Eq.~(\ref{qq'-logarithm}) can be generalized into three parameters as%
\footnote{We adopt the notation $\exp_q(x)$ in some of the equations of 
this section for aesthetical reasons.}:
\begin{equation}
 \ln_{q_1,q_2,q_3}x := \ln_{q_3}\exp\left({\ln_{q_1,q_2}x}\right).
\end{equation}
The procedure can be extended to an arbitrary number $n$ of parameters:
let us consider a set of $n$ parameters $\{q_1, q_2, \ldots, q_n\}$,
symbolically represented by $\{q_n\}$,
and let us define the $n$-parameter generalization of the logarithm
function in terms of the $(n-1)$-parameter logarithm
\begin{equation}
 \label{multi-log}
 \ln_{\{q_n\}} x := \ln_{q_n} \exp\left({\ln_{\{q_{n-1}\}}x}\right),
 \quad n>1.
\end{equation}
Note that $\ln_{q_n} x$ is the mono-parametric $q$-logarithm with $q=q_n$.
The $n$-parameter generalization of the exponential function must be
consistently defined as the inverse function of Eq.~(\ref{multi-log}),
\begin{equation}
 \ln_{\{q_n\}}\exp_{\{q_n\}}(x) = \exp_{\{q_n\}}(\ln_{\{q_n\}}x) = x,
\end{equation}
and thus,
\begin{equation}
 \label{multi-exp}
 \exp_{\{q_n\}}(x) := \exp_{\{q_{n-1}\}}\left({\ln \exp_{q_n}(x)}\right).
\end{equation}

Now we can promptly apply the recipes given by Eq.s (\ref{recipe-0}),
(\ref{recipe-1}), and (\ref{recipe-2-sum})--(\ref{recipe-2-product}),
and define
\begin{equation}
 \label{multi-sum}
 x\oplus_{\{q_n\}}y :=
 \ln_{\{q_n\}}\left(\exp_{\{q_n\}}(x)\,\exp_{\{q_n\}}(y)\right),
\end{equation}
\begin{equation}
 \label{multi-product}
 x\otimes_{\{q_n\}}y :=
 \exp_{\{q_n\}}\left({\ln_{\{q_n\}}x\,+\,\ln_{\{q_n\}}y}\right).
\end{equation}
With these operators, Eq.s~(\ref{lnq(xy)})--(\ref{eq(x+y)}) are
generalized into:
\begin{equation}
 \label{multiln(xy)}
 \ln_{\{q_n\}} (xy) = \ln_{\{q_n\}} x \oplus_{\{q_n\}} \ln_{\{q_n\}} y,
\end{equation}
\begin{equation}
 \label{multiln(xoqny)}
 \ln_{\{q_n\}} (x \otimes_{\{q_n\}} y) = \ln_{\{q_n\}} x + \ln_{\{q_n\}} y,
\end{equation}
\begin{equation}
 \label{multiexp(xoqny)}
 e_{\{q_n\}}^x e_{\{q_n\}}^y = e_{\{q_n\}}^{x \oplus_{\{q_n\}} y},
\end{equation}
\begin{equation}
 \label{multiexp(x+y)}
 e_{\{q_n\}}^{x+y} = e_{\{q_n\}}^x \otimes_{\{q_n\}} e_{\{q_n\}}^y.
\end{equation}
The motivation of Ref.~\cite{schwammle-ct} is still preserved:
the generalizations of Eq.~(\ref{multiln(xy)})--(\ref{multiln(xoqny)})
and
Eq.~(\ref{multiexp(xoqny)})--(\ref{multiexp(x+y)}) are:
\begin{equation}
 \ln_{\{q_{m+n}\}} (x \otimes_{q_1,\ldots,q_m} y) = \ln_{\{q_{m+n}\}} x 
 \oplus_{q_{m+1},\ldots,q_{m+n}} \ln_{\{q_{m+n}\}} y,
\end{equation}
\begin{equation}
 \exp_{\{q_{m+n}\}}(x\oplus_{q_{m+1},\ldots q_{m+n}}y) = 
 \exp_{\{q_{m+n}\}}(x)\otimes_{q_1,\ldots,q_m}\exp_{\{q_{m+n}\}}(y).
\end{equation}
In the former equations,
$\{q_{m+n}\}\equiv\{q_1,\ldots,q_m, q_{m+1},\ldots,q_n\}$,
and the order of the parameters cannot be changed.

It is straightforward to define the $\{q_n\}$-inverse operators
$\{q_n\}$-difference, $\{q_n\}$-ratio,
as well as the $\{q_n\}$-power operator; we shall not explicit
these relations here for the sake of brevity.

Analogously we can define multi-parametric $\{q_n\}$-dot-product
by applying the recipe suggested by Eq.~(\ref{recipe-qq'-dot}):
\begin{equation}
 \label{multi-dot}
 a\odot_{\{q_n\}}x := \ln_{\{q_n\}}\left[\exp_{\{q_n\}}(x)\right]^a.
\end{equation}

Properties of $\odot_{\{q_n\}}$ are straightforwardly analogous
to those of $\odot_{q,q'}$:
 \begin{equation}
 a\odot_{\{q_n\}}x = \ln_{q_n}e^{a\odot_{\{q_{n-1}\}}\ln e_{q_n}^x},
 \end{equation}
 \begin{equation}
 a\odot_{\{q_{n+m}\}}x = 
 \ln_{q_{n+1},\ldots,q_{n+m}}
     e^{a\odot_{q_1,\ldots,q_n}\ln e_{q_{n+1},\ldots,q_{n+m}}^x},
 \end{equation}
 \begin{equation}
 e_{\{q_{n+m}\}}^x =
 \left[e_{q_1,\ldots,q_n}%
        ^{a\odot_{q_1,\ldots,q_n}
         \ln e_{q_{n+1},\ldots,q_{n+m}}^x}\right]^{\frac{1}{a}},
 \end{equation}
 \begin{equation}
  a\odot_{\{q_{n}\}}(x\oplus_{\{q_{n}\}}y) =
  (a\odot_{\{q_{n}\}}x)\oplus_{\{q_{n}\}}(a\odot_{\{q_{n}\}}y),
 \end{equation}
 \begin{equation}
  (a+b) \odot_{\{q_{n}\}} x = 
  (a\odot_{\{q_{n}\}}x) \oplus_{\{q_{n}\}} (b\odot_{\{q_{n}\}}x).
 \end{equation}

\section{Conclusions and Perspectives}
\label{sec:conclusions}

We have introduced a two-parameter ($q,q'$)-algebraic structure, that naturally
emerges from the recently defined ($q,q'$)-logarithm and ($q,q'$)-exponential
\cite{schwammle-ct},
and we have investigated some of its properties.
The developments follows the lines of Ref.~ \cite{wang:q-algebra,epb:q-algebra},
in which it was introduced the mono-parametric $q$-algebra.
We have also introduced the ($q,q'$)-dot product, defined by repeated
($q,q'$)-sums.
Besides, we have extended the procedure, and defined multi-parametric
logarithm and exponential functions, and their corresponding algebras.

The possible applications of these developments are not still clear,
but we recall the central role played by the $q$-product in the
generalization of the central limit theorem and the $q$-Fourier transform
\cite{umarov-ct-stan:qclt-1,umarov-ct-mgm-stan:qclt-1,umarov-ct-mgm-stan-2,
umarov-ct:inverse-fourier,umarov-silvio:qclt};
it remains as a possibility the use of ($q,q'$)-product for further
developments of these theories.
We also recall that in the developments of the generalization of the
central limit theorem, it appears not only one, but a sequence of
$q_n$ parameters \cite{umarov-ct-stan:qclt-1}
(or $q_{\alpha,n}$ parameters, as in \cite{umarov-ct-mgm-stan-2}).
The existence of a $q$-triplet ($q_{sen}$, $q_{rel}$, $q_{stat}$)
was conjectured in \cite{ct:q-triplet-2004}, and it was observed in the solar wind
at the distant heliosphere \cite{burlaga-vinas:q-triplet}.
The connections of the theories that use multiple entropic indexes with the
developments here introduced is a possibility to be further investigated.

\section*{Acknowledgments}

This work is partially supported by CNPq -- Conselho Nacional de
Desenvolvimento Cient\'{\i}fico e Tecnol\'ogico (Brazilian Agency).

\section*{Appendix}

\renewcommand{\theequation}{A.\arabic{equation}}
\setcounter{equation}{0}

Here we present explicit (and unfortunately cumbersome) forms
of some expressions introduced in this paper.

\begin{description}

\item [($q,q'$)-logarithm] (Eq.~(\ref{qq'-logarithm})):
\begin{equation}
 \label{app-log}
 \ln_{q,q^{\prime}}x =
 \frac{1}{1-q^{\prime}}
  \left\{
         \exp\left[\frac{1-q^{\prime}}{1-q}\left(x^{1-q}-1\right)\right]-1
  \right\},
  \quad  x>0.
\end{equation}

\item [($q,q'$)-exponential] (Eq.~(\ref{qq'-exponential})):
\begin{equation}
 \label{app-exp}
 e_{q,q^{\prime}}^x =
 \left\{1+ \frac{1-q}{1-q^{\prime}} \ln \bigl[1+ (1-q^{\prime}) x  \bigr]
 \right\}_+^{\frac{1}{1-q}},
 \quad [1+(1-q^{\prime})x] > 0.
\end{equation}

\item [($q,q'$)-sum] (Eq.~(\ref{recipe-2-sum})):
\begin{eqnarray}
 \label{app-sum}
 x \oplus_{q,q'}y  =
 \frac{
       \omega_{q'}(x) \omega_{q'}(y)
       \exp\left\{
                  \left(\frac{1 - q}{1-q'}\right) 
                  \ln[\omega_{q'}(x)] \ln[\omega_{q'}(y)]
           \right\} -1
     }
     {1 - q'}, 
\end{eqnarray}
with $\omega_{q'}(u) \equiv 1+(1-q')u$, $\omega_{q'}(u)>0$.

\item [($q,q'$)-product] (Eq.~(\ref{recipe-2-product})):
\begin{equation}
 \label{app-product}
 x \otimes_{q,q'}y =
 \textstyle
 \left\{
   1 + \left( \frac{1-q}{1-q'} \right) 
       \ln \left[
{\scriptstyle
           {\left(e^{\frac{x^{1 - q}-1}{1 - q}} \right) }^{1 - q'}
           +
	   {\left(e^{\frac{y^{1 - q}-1}{1 - q}} \right) }^{1 - q'}
}
           -1
           \right]
 \right\}_+^{\frac{1}{1 - q}},
\end{equation}
with $x>0$, $y>0$, 
$\left[
       {(e^{\frac{x^{1 - q}-1}{1 - q}})}^{1 - q'}
      +{(e^{\frac{y^{1 - q}-1}{1 - q}})}^{1 - q'}
      -1
\right]
>0$.

\item [($q,q'$)-opposite] (Eq.~(\ref{qq'-opposite})):
\begin{equation}
 \label{app-opposite}
 \ominus_{q,q'}y = 
 \frac{
       \exp\left\{
                  \frac{-\ln[1+(1-q')y]}
                       {1+\left(\frac{1-q}{1-q'}\right)\ln[1+(1-q')y]}
           \right\}
           -1
      }
      {1-q'},
\end{equation}
with $[1+(1-q')y]>0$, 
$y\ne \frac{e^{\frac{1-q'}{q-1}}-1}{1-q'}$.

\item [($q,q'$)-difference] (Eq.~(\ref{qq'-difference})):
\begin{equation}
 \label{app-difference}
 x\ominus_{q,q'}y = 
 \frac{\exp\left\{
                  \frac{\ln\left[\frac{1+(1-q')x}{1+(1-q')y}\right]}
                       {1+\left(\frac{1-q}{1-q'}\right)\ln[1+(1-q')y]}
           \right\}-1
       }
       {1-q'},
\end{equation}
with $[1+(1-q')x]>0$, $[1+(1-q')y]>0$, 
$y\ne \frac{e^{\frac{1-q'}{q-1}}-1}{1-q'}$.

\item [($q,q'$)-inverse] (Eq.~(\ref{qq'-inverse})):
\begin{equation}
 \label{app-inverse}
 1 \oslash_{q,q'}y = 
 \left\{
        1 + \left( \frac{1-q}{1-q'} \right) 
        \ln \left[
                  2-{\left(e^{\frac{y^{1 - q}-1}{1 - q}} \right)}^{1 - q'}
            \right]
 \right\}_+^{\frac{1}{1 - q}},
\end{equation}
with $0<y<\left[1+\frac{1-q}{1-q'}\ln 2 \right]^{\frac{1}{1-q}}$
for $q'<1$, 
and $y>\left[1+\frac{1-q}{1-q'}\ln 2 \right]^{\frac{1}{1-q}}$
for $q'>1$.

\item [($q,q'$)-ratio] (Eq.~(\ref{qq'-ratio})):
\begin{equation}
 \label{app-ratio}
 x \oslash_{q,q'}y=
 \textstyle
 \left\{
        1 + \left( \frac{1-q}{1-q'} \right) 
	\ln \left[
{\scriptstyle
                  {\left(e^{\frac{x^{1 - q}-1}{1 - q}} \right)}^{1 - q'}
                 -{\left(e^{\frac{y^{1 - q}-1}{1 - q}} \right)}^{1 - q'}
}
                 +1 
            \right]
 \right\}_+^{\frac{1}{1 - q}},
\end{equation}
with $x>0$, $y>0$, 
$ \left[
        {\left(e^{\frac{x^{1 - q}-1}{1 - q}} \right)}^{1 - q'}
        -{\left(e^{\frac{y^{1 - q}-1}{1 - q}} \right)}^{1 - q'}
        +1 
  \right] > 0$.

\item [($q,q'$)-dot-product] (Eq.~(\ref{recipe-qq'-dot})):
\begin{equation}
 \label{app-dot}
 a\odot_{q,q'}x=
 \frac{%
       \exp
        \left\{\left(\frac{1-q'}{1-q}\right)
          \left[
{\scriptstyle
	   \left[1+\left(\frac{1-q}{1-q'}\right)
	   \ln\left(1+(1-q')x\right)
	   \right]_+^a -1
}
	  \right]
        \right\}-1
       }
      {1-q'},
\end{equation}
with $[1+(1-q')x]>0$.

\end{description}


\end{document}